\documentclass[leqno]{amsart}
\usepackage{amsmath}
\usepackage{amssymb}
\usepackage{amsthm}
\usepackage{enumerate}
\usepackage[mathscr]{eucal}
\usepackage{graphicx}
\usepackage{hyperref}
\hypersetup{colorlinks,linkcolor={blue},citecolor={blue},urlcolor={red}}
\usepackage{color}
\theoremstyle{plain}

\newtheorem{lemma}{Lemma}[section]

\theoremstyle{definition}

\newtheorem{remark}{Remark}[section]

\newtheorem{example}{Example}

\usepackage[pagewise]{lineno}
\makeatletter
\@namedef{subjclassname@2020}{%
\textup{2020} Mathematics Subject Classification}
\makeatother
\begin{document}	

\title[ LCD code-based Multi-secret sharing scheme]{ Linear complementary dual code-based Multi-secret sharing scheme}
\author[Ghosh, Bhowmick, Maurya, Bagchi]{Haradhan Ghosh, Sanjit Bhowmick, Pramod Kumar Maurya, Satya Bagchi}
\newcommand{\acr}{\newline\indent}

\address[Ghosh]{Department of Mathematics\\ National Institute of Technology Durgapur\\ Durgapur 713209\\ West Bengal\\ INDIA}
\email{hg.20ma1104@phd.nitdgp.ac.in}

\address[Bhowmick]{Department of Mathematics\\ National Institute of Technology Durgapur\\ Durgapur 713209\\ West Bengal\\ INDIA}
\email{bhowmick392@gmail.com}
\address[Maurya]{School of Computer Science and Engineering\\ Vellore Institute of Technology\\ Vellore 632014\\ INDIA}
\email{pramodkumar.maurya@vit.ac.in}
\address[Bagchi]{Department of Mathematics\\ National Institute of Technology Durgapur\\ Durgapur 713209\\ West Bengal\\ INDIA}
\email{satya.bagchi@maths.nitdgp.ac.in}

\thanks{The first author of the paper would like to thank  CSIR-HRDG, India, for support financially to carry out this work. This work is also supported by the National Board of Higher Mathematics (NBHM), Government of India (Grant No. 02011/2/2019 NBHM (R.P)/R\& D II/1092).}

\subjclass[2020]{Primary 94A62, 94B05, 13H05}
	\keywords{Secret sharing scheme, Multi-secret sharing scheme, Local ring, Linear complementary dual codes}

\begin{abstract}
Hiding a secret is needed in many situations. Secret sharing plays an important role in protecting information from getting lost, stolen, or destroyed and has been applicable in recent years. A secret sharing scheme is a cryptographic protocol in which a dealer divides the secret into several pieces of share and one share is given to each participant. To recover the secret, the dealer requires a subset of participants called access structure. In this paper, we present a multi-secret sharing scheme over a local ring based on linear complementary dual codes using Blakley's method. We take a large secret space over a local ring that is greater than other code-based schemes and obtain a perfect and almost ideal scheme.
\end{abstract}

\maketitle
\newcommand{\thedate}{\today}
%\thedate
\section{Introduction} 

Secret sharing is a method for distributing a secret among a group of participants, each of whom is allocated a share of the secret in which some subgroups of them can recover the secret. It is one of the most important cryptographic protocols. Secret sharing scheme (SSS) was first introduced by Blakley \cite{blakly} and Shamir \cite{shamir} independently in the late 1970s. Working principles of these schemes are based on protecting the encryption keys and apply for $k$ out of $n$ threshold access structures that includes all subsets of cardinality at least $k$ for $1\leq k < n$, and it is denoted as $(k,n)$ threshold scheme. For an $(k,n)$ threshold scheme, $k$ or more participants can reconstruct the secret, but $(k-1)$ or fewer participants can not recover the secret. SSS is accomplished in information security, information hiding, threshold cryptography, access control and many others.\\

Multi-secret sharing scheme (MSSS) is an important family of SSSs. It is a case in which many secrets need to be shared. In other words, a multi-secret sharing scheme is a protocol to share $m$ arbitrarily related secrets $s_1$, $s_2$, $\dots$, $s_m$ among a set of participants $\mathcal{P}$. If ($A_1$, $A_2$, $\dots$, $A_m$) be an $m$-tuple of monotone access structures on $\mathcal{P}$ then any subset of $\mathcal{P}$, enables to recover $s_i$, can compute $s_i$. Whereas any subset of $\mathcal{P}$, which is not enable to recover $s_i$ even if knowing some other information about $s_i$, cannot recover the secrets.\\

In 1994, He and Dawson \cite{he} proposed an MSSS scheme based on a one-way function. In that scheme, many secrets are reconstructed stage-by-stage in the dealer's predetermined order, and every participant keeps only one secret share. Later, Harn \cite{harn} improved the He-Dawson \cite{he} scheme to reduce the total number of shared values. In 2005, Pang et al. \cite{pang} proposed a MSSS using an $(n+p-1)$-th degree Lagrange interpolation polynomial. In their scheme, the degree of the polynomial is dynamic, for this  Lagrange interpolation operation becomes more complex, and also, computing time and storage requirement are large. Later, Li et al. \cite{huixian} proposed a MSSS using a fixed degree Lagrange interpolation polynomial. They have shown the scheme needs less storage demand than Pang et al. \cite{pang} scheme.
Bai \cite{bai} presented a reliable image secret sharing method that incorporates two k-out-of-n secret sharing scheme using matrix projection SSS and Shamir's \cite{shamir} SSS. In 2008, Ranhua et al. \cite{shi} proposed a new threshold MSSS in which the secret can be reconstructed by solving a system of linear equations.
\medskip

An essential class of secret sharing schemes is those which are based on linear codes. The property of linear complementary dual (LCD) codes was studied by Massey \cite{masey} and he constructed some LCD codes over finite fields. Mutto et al. \cite{mutto} analyzed reversible codes over $\mathbb{F}_q$. In 2013, Chen et al. \cite{CHEN} presented a strongly multiplicative linear ramp SSS based on error-correcting codes. Later, Herranz et al. \cite{her} proposed an MSSS where each share has constant length, and also they prove the computational cost by random oracle model. In 2015, Calkavur et al. \cite{calkavur} considered an MSSS based on error-correcting codes with a bounded distance algorithm. Xing et al. \cite{xing} proposed a fair $(t,n)$ threshold SSS using two variable one-way hash function, also they use Lagrange interpolation polynomial and polynomial verification to detect whether there is a dishonest participant in the system. After that Kabirirad et al. \cite{KABIR} presented a $(t,n)$ $(t\neq 2)$ multi-secret image sharing based on Boolean operations and they provide experiments to show that their scheme is robust against statistical and differential attacks. In 2018, Molla et al. \cite{molla} developed a new approach to construct a SSS based on field extensions. Later, Calkavur et al. \cite{cal} construct an image SSS based on Shamir \cite{shamir} secret sharing. Recently Alahmadi et al. \cite{article} presented a multi-secret sharing scheme on LCD code $C$ with length $n$ over finite field $\mathbb{F}_q$. In the scheme, the secret key is chosen from $\mathbb{F}_q^k$, which is a limitation of the scheme, where $k$ is the dimension of $C$. \\

After reviewing the previous work, we present a multi-secret sharing scheme based on LCD codes over a local ring $R$. In this work, we chose the secret key from $R^n$ and shows that the scheme is more secure than Alahmadi et al. scheme, where $n$ is the length of the code $C$. To recover the secret, we use the property of LCD codes. We also make a comparison between the proposed scheme and some other code-based schemes.\\

The paper organised as follows. In Section 2, some preliminaries about local ring, LCD codes and some definitions are given. In Section 3, we present a new multi-secret sharing scheme. In Section 4, some examples are given to verify our scheme. We analyze our scheme in Section 5. We compare our scheme with other existing schemes in Section 6. Finally, we conclude our work in Section 7.

 \section{Preliminaries}

A finite commutative ring $R$ is called a local ring if it has a unique maximal ideal. All fields are local rings. The ring $\mathbb{Z}/p^i\mathbb{Z}$ is local, for prime $p$ and $i\in \mathbb{N}$. In a local, every element is either a unit or a nilpotent.\\

A linear code $C$ over a local ring $R$ is an $R$-submodule of $R^n$, where $n$ is a positive integer. A linear code $C$ is called an LCD code if $C\cap C^\perp$=$\{0\}$, where $C^\perp$ is the dual code of $C$. If $C$ is LCD then so is $C^\perp$.\\
Throughout this paper, we consider $R$ as a commutative local ring. The following results are well known from \cite{sanjit}.
\begin{lemma}
Let $A$ be an $k\times n$ matrix over $R$, then $A$ is full row-rank if and only if $A$ is right invertible. 
\end{lemma}
\begin{lemma}
Let $A$ be a matrix over $R$, then the following are equivalent-
\begin{enumerate}[1.]
    \item $A$ is invertible.
    \item $A$ is non-singular.
    \item $A$ is full row-rank.
\end{enumerate}
\end{lemma}

\begin{lemma}
Let $A$ be an $k\times n$ matrix over $R$, then $A$ is not full row-rank if and only if there exists a non-zero vector $x$ in $R^k$ such that $Ax^T=0$.
\end{lemma}

We have the following result from the above theorem, which plays an important role in our paper.
\begin{lemma}
Let $C$  be a linear code over $R$ with a generator matrix $G$ and a parity-check matrix $H$, then $C$ is an LCD code if and only if 
$\begin{pmatrix} G\\ H \end{pmatrix}$ is invertible.
\end{lemma}
\medskip
\subsection{Terminologies}
Some terminology, we need to define a secret sharing scheme:
\begin{enumerate}[i.]
    \item Shares- Shares are part of information. In a secret sharing scheme, these shares can recover the secret, and shares must be kept highly confidential.
    \item Share set- The set of all possible shares of a secret.
    \item Participants- The parties that receive the share of a secret.
    \item Dealer- Dealer picks the secret and distributes the shares among participants.
    \item Access structure- Access structure is the set of minimal coalition sets, the elements of this set are the authorized combinations of participants whose share can use to reconstruct the secret. In notation, let $\mathcal{P}$ be the set of participants. A collection $\mathcal{A}$ $\subseteq$ $2^\mathcal{P}$ is monotone if $B \in \mathcal{A}$ and $B \subseteq C$ implies $C \in$  $\mathcal{A}$. An access structure is a monotone collection $\mathcal{A}$ $\subseteq$ $2^\mathcal{P}$ of nonempty subsets of $\mathcal{P}$ and sets in $\mathcal{A}$ are authorized.
    
\end{enumerate}
\noindent
A secret sharing scheme involves a dealer who has a secret, a set of parties, and an access structure $\mathcal{A}$. A secret sharing scheme for $\mathcal{A}$ is a method by which the dealer distributes shares to the parties such that any subset in $\mathcal{A}$ can reconstruct the secret from its shares, while any unauthorized subset cannot reveal any information on the secret.\\ 
\medskip
\section{Multi-secret sharing scheme based on LCD codes}
We present a new multi-secret sharing scheme based on LCD codes over a local ring. We use Blakley's \cite{blakly} method to explain our approach.

\subsection{Scheme description}
Let $C$ be an $[n,k]$ LCD code over a local ring $R$ with a generator matrix $G$ of order $k\times n$, where $\left \lceil{\frac{n}{2}}\right \rceil  \leq k \leq \left \lceil{\frac{n}{2}}\right \rceil +r$, $0\leq r\leq \left \lfloor{\frac{n}{2}}\right \rfloor -1$.
Let $R^n$ be the secret space and $s=s_1s_2 \cdots s_n \in R^n$ be the secret.
\subsubsection{Secret distribution}
A dealer fixed a code $C$ of dimension $k$ with a generator matrix $G$ and a parity-check matrix $H$ such that $GH^T=0$.
The dealer chooses $l_i\in R^k$ and calculate $c_i=l_i G$.\\
Since $C$ is an LCD code therefore $C\cap C^\perp=\{0\}$. Taking first $(n-k)$ positions of $l_i$ as $l'_i$, and then calculate $c'_i=l'_iH$.
Calculating $x_i=c_is^T$ and $y_i=c'_is^T$, where $s^T$ is the transpose of $s$, the dealer distributes to a participant $P_i=(c_i,x_i,y_i)$.
\subsubsection{Secret recover}
The dealer needs at least $k$ participants whose $c_i$'s are linearly independent to recover the secret. Let a set of participants be $P_1, P_2,\dots, P_k$.
So, the dealer gets the values $c_1,c_2, \dots, c_k$, $x_1,x_2, \dots, x_k$, $y_1,y_2, \dots, y_{k-1}$ and $y_k$.\\

The dealer calculates $l''_i$ such that $l''_iG=c_i$ for $i=1,2,...,k$. Let $L=
\begin{pmatrix}
l''_1\\
\vdots \\
l''_k
\end{pmatrix}.$ Then the order of $L$ is $k\times k$ and the rank is $k$ because each $l''_i$ is linearly independent. After deleting last $2r$ entries from each $l''_i$, we have a matrix $L'$ of order $k\times n-k$. We know that the rank of a matrix $A$ is the highest order non-vanishing minor of $A$. Since the rank of $L$ is $k$ and the order of $L'$ is $k\times n-k$, therefore we have a non-vanishing minor of $L'$ of order $n-k$. So, we have linearly independent $l'''_i$ (say) from $R^{n-k}$ for $i=1,2,\dots,n-k$.\\
\noindent
Next calculate $c''_i=l'''_iH$, $i=1,\dots,n-k$, where each $l'''_i$ are the $i$th row of $L'$. So, rank of $L'$ is $n-k$ and hence each $c''_i$ are linearly independent.\\
Let $A=\begin{pmatrix} A_1\\ A_2 \end{pmatrix},$ where $A_1=\begin{pmatrix} c_1\\ \vdots \\ c_k \end{pmatrix}$ and $A_2=\begin{pmatrix}
c''_1\\
\vdots \\
c''_{n-k}
\end{pmatrix}.$ The rank of both $H$ and $L'$ are $n-k$. By Sylvester theorem \cite{zhang} rank of $A_2$ is $n-k$. All $c_i\in C$ are linearly independent and all $c''_i\in C^\perp$ are also linearly independent. Therefore, the rank of $A$ is $k+(n-k)=n$.\\

Let $B=(x_1, \dots, x_k, y_1, \dots, y_k)^T$. From $B$, taking all $x_i$ and such $y_j$ for which $c''_j$ are linearly independent, where $j=1,2,\dots,n-k$. Therefore, $B':=(x_1, \dots, x_k, y_1, \dots, y_{n-k})^T$.\\
Now, consider the system of linear equations $$As^T=B',$$ where $s=(s_1,s_2,...,s_n)$ is the secret. Since rank of $A$ is $n$, we have an unique solution of the above system of equations.
Therefore the secret $s$ is recovered.
\begin{remark}
For $n$ is even and $r=0$, i.e., for $k=\frac{n}{2}$, the scheme is easy to check, because in this case the order of $G$ and $H$ are same and it is $k\times n$. 
\end{remark}
\section{Examples}
\begin{example}
Let $C$ be an $[8,4]$ LCD code over $\mathbb{Z}_4$ with a generator matrix \[G=
\begin{pmatrix}
1&0&0&0&0&1&2&1\\
0&1&0&0&1&2&3&1\\
0&0&1&0&0&0&3&2\\
0&0&0&1&2&3&1&1
\end{pmatrix}\] 

and a parity-check matrix $H=$
$\begin{pmatrix}
0&3&0&2&1&0&0&0\\
3&2&0&1&0&1&0&0\\
2&1&1&3&0&0&1&0\\
3&3&2&3&0&0&0&1
\end{pmatrix}$.
Let $s=22000000$ be the secret.\\
Consider codewords  $l_i$'s as, $l_1=1000$, $l_2=0010$, $l_3=1100$, $l_4=1110$, $l_5=2000$, $l_6=0200$, $l_7=3000$, $l_8=1200$, $l_9=1122$, $l_{10}=2200$ and so on.\\

\noindent \textbf{Dealer's calculation:}
The dealer calculates $c_i=l_iG$ as
$c_1=10000121$, $c_2=00100032$, $c_3=11001312$, $c_4=11101300$, $c_5=20000202$, $c_6=02002022$, $c_7=30000323$, $c_8=12002103$, $c_9=11221110$, $c_{10}=22002220$, and so on, and $c'_i=l_iH$ as $c'_1=03021000$, $c'_2=21130010$, $c'_3=31031100$, $c'_4=12121110$, $c'_5=02002000$, $c'_6=20020200$, $c'_7=01023000$, $c'_8=23001200$, $c'_9=11231122$, $c'_{10}=22022200$, and so on.
Calculating
$x_i=c_is^T$ as $x_1=2$, $x_2=0$, $x_3=0$, $x_4=0$, $x_5=0$, $x_6=0$, $x_7=2$, $x_8=2$, $x_9=0$, $x_{10}=0$, and so on, and $y_i=c'_is^T$ as $y_1=2$, $y_2=2$, $y_3=0$, 
$y_4=2$, $y_5=0$, $y_6=0$, $y_7=2$, $y_8=2$, $y_9=0$, $y_{10}=0$, and so on. The dealer chooses the shares of each participants as $P_i=(c_i,x_i,y_i)$.\\

\noindent \textbf{Dealer's recover:}
To recover the secret the dealer needs at least $k=4$ independent shares.
Suppose independent participant are $P_1$, $P_3$, $P_4$ and $P_9$.
Let \[G'=
\begin{pmatrix}
c_1\\
c_3\\
c_4\\
c_9\\
\end{pmatrix}
=
\begin{pmatrix}
1&0&0&0&0&1&2&1\\
1&1&0&0&1&3&1&2\\
1&1&1&0&1&3&0&0\\
1&1&2&2&1&1&1&0\\
\end{pmatrix}.\]
Find $l'_1$ such that $l'_1G=c_1$. If $l'_1=abcd$ then $l'_1G=c_1$ gives  $a=1$, $b=0$, $c=0$, $d=0$, $b+2d=0$, $a+2b+3d=1$, $2a+3b+3c+d=2$, and $a+b+2c+d=1$.
By solving this equations we have $a=1$, $b=0$, $c=0$, $d=0$.
So $l'_1=1000$.\\

Similarly solving $l'_3G=c_3$, $l'_4G=c_4$ and $l'_9G=c_9$, we have $l'_3=1100$, $l'_4=1110$ and $l'_9=1122$ respectively.

\noindent
Next calculate  $c''_i=l'_iH$ for $i=1$, $3$, $4$, $9$. We get
$c''_1=03021000$, $c''_3=31031100$, $c''_4=12121110$, $c''_9=11231122$.\\
Take $H'=$
$\begin{pmatrix}
c''_1\\
c''_3\\
c''_4\\
c''_9
\end{pmatrix}$
=
$\begin{pmatrix}
0&3&0&2&1&0&0&0\\
3&1&0&3&1&1&0&0\\
1&2&1&2&1&1&1&0\\
1&1&2&3&1&1&2&2
\end{pmatrix}$.

Then
$\begin{pmatrix}
G'\\
H'
\end{pmatrix}$
=
$\begin{pmatrix}
1&0&0&0&0&1&2&1\\
1&1&0&0&1&3&1&2\\
1&1&1&0&1&3&0&0\\
1&1&2&2&1&1&1&0\\
0&3&0&2&1&0&0&0\\
3&1&0&3&1&1&0&0\\
1&2&1&2&1&1&1&0\\
1&1&2&3&1&1&2&2
\end{pmatrix}$.

Let $s=s_1s_2s_3s_4s_5s_6s_7s_8$ be the secret.
Consider the system of equations $\begin{pmatrix}
G'\\
H'
\end{pmatrix}s^T=$ $(x_1x_3x_4x_9y_1y_3y_4y_9)^T=(20002020)^T$.
Since the rank of $A$ is $8$, the solution to the system of equations is unique.
By solving this system of equations we get $s=22000000$.

\end{example}

\begin{example}
Let $C$ be an $[8,5]$ LCD code over $\mathbb{F}_2$ with generator matrix \[G=
\begin{pmatrix}
1&0&0&0&0&0&0&0\\
0&1&0&0&0&1&0&0\\
0&0&1&0&0&0&1&1\\
0&0&0&1&0&0&1&0\\
0&0&0&0&1&1&1&1\\
\end{pmatrix}\]
and the parity-check matrix $H=$
$\begin{pmatrix}
0&1&0&0&1&1&0&0\\
0&0&1&1&1&0&1&0\\
0&0&1&0&1&0&0&1\\
\end{pmatrix}$.
Let the secret be $s=11000000\in \mathbb{F}_2^8$.
So, here $n=8$, $k=5$ and $r=1$.\\
Consider codewords $l_i$'s as, $l_1=10000$, $l_2=01000$, $l_3=00100$, $l_4=00010$, $l_5=00001$, $l_6=11000$, $l_7=10100$, $l_8=10010$, $l_9=10001$, $l_{10}=01100$ and so on.\\

\noindent \textbf{Dealer's calculation:}
The dealer calculate $c_i=l_iG$ as
$c_1=10000000$, $c_2=01000100$, $c_3=00100011$, $c_4=00010010$, $c_5=00001111$,  $c_6=110001000$, $c_7=10100011$, $c_8=10010010$, $c_9=10001111$, $c_{10}=01100111$ and so on. 
\noindent
Deleting last $2r(=2)$ entries from each $l_i$ and then calculate $c'_i=l_iH$, we have $c'_1=01001100$, $c'_2=00111010$, $c'_3=00101001$, $c'_4=00000000$, $c'_5=00000000$, $c'_6=01011110$, $c'_7=01100101$, $c'_8=01001100$, $c'_9=01001100$, $c'_{10}=00010011$, and so on.\\
Then calculate
$x_i=c_is^T$ as $x_1=1$, $x_2=1$, $x_3=0$, $x_4=0$, $x_5=0$, $x_6=0$, $x_7=1$, $x_8=1$, $x_9=1$, $x_{10}=1$, and so on, and $y_i=c'_is^T$ as $y_1=1$, $y_2=0$, $y_3=0$, $y_4=0$, $y_5=0$, $y_6=1$, $y_7=1$, $y_8=1$, $y_9=1$, $y_{10}=0$, and so on.
\noindent The dealer choose the shares of each participants as $P_i=(c_i,x_i,y_i)$.\\

\noindent \textbf{Dealer's recover:}
To recover the secret the dealer needs at least $k=5$ shares. Let the participants $P_1,P_3,P_4,P_6,P_9$ give their shares to recover the secret.
Let $G'=$
$\begin{pmatrix}
c_1\\
c_3\\
c_4\\
c_6\\
c_9
\end{pmatrix}$ 
=
$\begin{pmatrix}
1&0&0&0&0&0&0&0\\
0&0&1&0&0&0&1&1\\
0&0&0&1&0&0&1&0\\
1&1&0&0&0&1&0&0\\
1&0&0&0&1&1&1&1\\
\end{pmatrix}$.\\
Find $l'_1$ such that $l'_1G=c_1$. If $l'_1=abcde$ then $l'_1G=c_1$ gives $a=1$, $b=0$, $c=0$, $d=0$, $e=0$, $b+e=0$, $c+d+e=0$, $c+e=0$. By solving this equations we have $a=1$, $b=0$, $c=0$, $d=0$, $e=0$.
So $l'_1=10000$.\\
Similarly solving $l'_3G=c_3$, $l'_4G=c_4$, $l'_6G=c_6$ and $l'_9G=c_9$ we have $l'_3=00100$, $l'_4=00010$, $l'_6=11000$ and $l'_9=10001$ respectively.\\
After deleting last 2 entries from each $l'_i$ $(i=1,3,4,6,7)$, calculate $c''_i=l'_iH$ for $i=1,3,4,6,7$.
we get $c''_1=01001100$, $c''_3=00101001$, $c''_4=00000000$, $c''_6=01110110$, and $c''_9=01001100$.
Take 3 linearly independent $c''_i$ as $c''_1$, $c''_3$ and $c''_6$.\\
Let $A=
\begin{pmatrix}
c_1\\ c_3\\c_4\\c_6\\
c_9\\
c''_1\\
c''_3\\
c''_6
\end{pmatrix}
=
\begin{pmatrix}
1&0&0&0&0&0&0&0\\
0&0&1&0&0&0&1&1\\
0&0&0&1&0&0&1&0\\
1&1&0&0&0&1&0&0\\
1&0&0&0&1&1&1&1\\
0&1&0&0&1&1&0&0\\
0&0&1&0&1&0&0&1\\
0&1&1&1&0&1&1&0
\end{pmatrix}.$
Also, let $s=(s_1s_2s_3s_4s_5s_6s_7s_8)$ be the secret.
Consider the system of equations $As^T=$ $(x_1x_3x_4x_6x_9y_1y_3y_6)=(10001101)^T.$
Since rank of $A$ is $8$, the solution to the system of equations is unique.
By solving this system of equations we get $s=11000000$.

\end{example}
\begin{example}
Let $C$ be $[8,4]$ LCD code with generator matrix $G=$
$\begin{pmatrix}
1&0&0&0&0&0&0&1\\
0&1&0&0&1&1&0&1\\
0&0&1&0&1&1&0&0\\
0&0&0&1&1&0&1&1
\end{pmatrix}$ and then the parity-check matrix $H=$
$\begin{pmatrix}
0&1&1&1&1&0&0&0\\
0&1&1&0&0&1&0&0\\
0&0&0&1&0&0&1&0\\
1&1&0&1&0&0&0&1
\end{pmatrix}$.
Let $s=11000001$ be the secret.\\
Consider codewords  $l_i$'s as, $l_1=1000$, $l_2=0100$, $l_3=0010$, $l_4=0001$, $l_5=1100$, $l_6=1010$, $l_7=1001$, $l_8=0110$, $l_9=0011$, $l_{10}=0101$, $l_{11}=1110$, $l_{12}=1101$, $l_{13}=0111$, $l_{14}=1011$, $l_{15}=1111$, $l_{16}=0000$.\\

\noindent \textbf{Dealer's calculation:}
The dealer calculate $c_i=l_iG$ as
$c_1=10000001$, $c_2=01001101$, $c_3=00101100$, $c_4=00011011$, $c_5=11001100$,  $c_6=101011010$, $c_7=10011010$, $c_8=01100001$, $c_9=00110111$, $c_{10}=01010110$, $c_{11}=11100000$, $c_{12}=11010110$, $c_{13}=01111010$, $c_{14}=10110110$, $c_{15}=11111011$, $c_{16}=00000000$\\
and 
$c'_i=l_iH$ as $c'_1=01111000$, $c'_2=01100100$, $c'_3=00010010$, $c'_4=11010001$, $c'_5=00011100$, $c'_6=01101010$, $c'_7=10101001$, $c'_8=01110110$, $c'_9=11000011$, $c'_{10}=10110101$, $c'_{11}=00001110$, $c'_{12}=11001101$, $c'_{13}=10100110$, $c'_{14}=10111011$, $c'_{15}=11011111$, and $c'_{16}=00000000$.\\
Then calculate $x_i=c_is^T$ as $x_1=0$, $x_2=0$, $x_3=0$, $x_4=1$, $x_5=0$, $x_6=0$, $x_7=1$, $x_8=0$, $x_9=1$, $x_{10}=1$, $x_{11}=0$, $x_{12}=0$, $x_{13}=1$, $x_{14}=1$, $x_{15}=1$, $x_{16}=0$\\
and $y_i=c'_is^T$ as $y_1=1$, $y_2=1$, $y_3=0$, $y_4=1$, $y_5=0$, $y_6=1$, $y_7=0$, $y_8=1$, $y_9=1$, $y_{10}=0$, $y_{11}=0$, $y_{12}=1$, $y_{13}=1$, $y_{14}=0$, $y_{15}=1$, $y_{16}=0$. The dealer choose the shares of each participants as $P_i=(c_i,x_i,y_i)$.\\

\noindent \textbf{Dealer's recover:}
To recover the secret the dealer needs at least $k=4$ independent shares.
Let the 4 independent participants are $P_1$, $P_5$, $P_{11}$ and $P_{15}$.
Let \[G'=
\begin{pmatrix}
c_1\\
c_5\\
c_{11}\\
c_{15}\\
\end{pmatrix} 
=
\begin{pmatrix}
1&0&0&0&0&0&0&1\\
1&1&0&0&1&1&0&0\\
1&1&1&0&0&0&0&0\\
1&1&1&1&1&0&1&1\\
\end{pmatrix}.\] Find $l'_1$ such that $l'_1G=c_1$. If $l'_1=abcd$ then $l'_1G=c_1$ gives  $a=1$, $b=0$, $c=0$, $d=0$, $b+c+d=0$, $d=0$, and $a+b+d=1$. By solving this equations we have $a=1$, $b=0$, $c=0$, $d=0$.
So $l'_1=1000$.

Similarly solving $l'_5G=c_5$, $l'_{11}G=c_{11}$ and $l'_{15}G=c_{15}$, we have $l'_5=1100$, $l'_{11}=1110$ and $l'{15}=1111$ respectively.
\noindent Next calculate  $c''_i=l'_iH$ for $i=1, 5, 11, 15$. We get
$c''_1=01111000$, $c''_5=00011100$, $c''_{11}=00001110$, $c''_{15}=11011111$.\\
Take $H'=$
$\begin{pmatrix}
c''_1\\
c''_5\\
c''_{11}\\
c''_{15}
\end{pmatrix}$
=
$\begin{pmatrix}
0&1&1&1&1&0&0&0\\
0&0&0&1&1&1&0&0\\
0&0&0&1&0&0&1&0\\
1&1&0&1&0&0&0&1
\end{pmatrix}$.
Let $A=$
$\begin{pmatrix}
G'\\
H'
\end{pmatrix}$
=
$\begin{pmatrix}
1&0&0&0&0&0&0&1\\
1&1&0&0&1&1&0&0\\
1&1&1&0&0&0&0&0\\
1&1&1&1&1&0&1&1\\
0&1&1&1&1&0&0&0\\
0&0&0&1&1&1&0&0\\
0&0&0&1&0&0&1&0\\
1&1&0&1&0&0&0&1

\end{pmatrix}$.

Let $s=(s_1s_2s_3s_4s_5s_6s_7s_8)$ be the secret.
Consider the system of equations $As^T=$ $(x_1x_5x_{11}x_{15}y_1y_5y_{11}y_{15})^T=(00011001)^T.$
Since rank of $A$ is $8$, the solution to the system of equations is unique.
By solving this system of equations we get $s=11000001$.
\end{example}

\section{Analysis and discussion}
\subsection{Security analysis} To recover the secret we need at least $k$ independent participants. Let us assume that $t$ independent participants together try to guess the secret where $t<k$. Then they have $t$ linearly independent codewords $c_i$, $i=1,2,\dots,t$. Let $V_t$ be the span of these $c_i$'s and since all $c_i$'s are linearly independent, $V_t$ is a subspace $C$ of dimension $t$ where $C$ is a vector space of dimension $k$ over $\mathbb{F}_q$. If $W_t$ is the complementary subspace of $V_t$ into $C$ then $C=V_t\oplus W_t$ and the dimension of $W_t$ is $(k-t)$. The number of times \cite{article} any basis of $V_t$ can be extended into a basis of $C$ is equal to 
$$ X(k,t)=\frac{\prod_{i=0}^{k-1}(q^k-q^i)}{(k-t)!\prod_{i=0}^{t-1}(q^t-q^i)}.$$
For a basis of $W_t$, there are $q^{k-t}$ choices for shares of the codewords of this basis. So the probability to get the secret of this $t$ independent participants is $$\frac{1}{X(k,t) q^{k-t}}.$$
For large values of $k$ and $q$ this probability is negligible.

\subsection{Information theoretic efficiency}
In a secret sharing scheme, the information rate $\rho$ of an access structure is the ratio between the size of the secret and the size of the largest share given to any participant. Since the size of the secret is $n$ and the size of the largest share is $n+2$ therefore $$\rho = \frac{n}{n+2}.$$
So for $n\to \infty$ the information rate $\rho \to 1$ and then the scheme becomes an ideal scheme.

\section{Comparison of schemes}
We compare our scheme with other code-based secret sharing schemes for an $[n,k]$ code $C$ over $\mathbb{F}_q$ by means of the number of participants $(A)$, the size of a secret $(B)$, number of coalitions $(C)$, and the information rate $\rho$.

\begin{table}[htp]
    \centering
    \begin{tabular}{|c|c|c|c|c|c|c|}
    \hline
    System  & Messey's  & Ding et al.  & Calkavur's &  Alahmadi et al.  & \multicolumn{2}{|c|}{This scheme} \\ \cline{6-7}
        &  \cite{masey}&  \cite{ding} &  \cite{calkavur}&   \cite{article} & over $\mathbb{F}_q$ & over $R$ \\
        
    \hline
     
      A & $n-1$ & $n$ & $n$ & $q^k$ & $q^k$ & $r^k$ \\
      B & $q$ & $q^k$ & $q^k$ & $q^k$ & $q^n$ & $r^n$\\
      C & $\binom{n}{k}$ & $\binom{n}{k}$ & $\geq {\binom{n}{d-t}}$ & $\frac{\prod_{i=0}^{k-1}(q^k-q^i)}{k!}$ & $\frac{\prod_{i=0}^{k-1}(q^k-q^i)}{k!}$ & $\frac{\prod_{i=0}^{k-1}(r^k-r^i)}{k!}$ \\
      $\rho$ & $1$ & $\frac{k}{{k-1}}$ & $1$  & $\frac{k}{{k+1}}$ & $\frac{n}{n+2}$&  $\frac{n}{n+2}$\\
      \hline
    \end{tabular}
    \caption{Comparison of schemes}
    \label{table 1.}
\end{table}

All the schemes in Table \ref{table 1.} are ideal because the size of each secret is equal to the size of any shares. Here $t$ is the error-correcting capacity of the code, and $r$ is the cardinality of the local ring $R$. Also, the scheme \cite{ding} is a construction based on linear algebra, and the scheme \cite{calkavur} is based on decoding. Our scheme and the scheme \cite{article} are based on Blakley's \cite{blakly} method and obtain a system of linear equations that has a unique solution. Hence, the secret will be recovered definitely. Since all finite fields are local rings, our scheme works on both $\mathbb{F}_q$ and $R$. If we take a local ring, it is more difficult for an adversary to do a random search.

\section{Conclusion}
In this paper, we have presented a new multi-secret sharing scheme based on LCD codes over a local ring. To improve the size of the secret space, we took a local ring instead of a field. We have determined the information rate which is close to one, and hence our scheme is ideal. For higher values of $k$, i.e., for high dimension, the security of our scheme stands well. Also, since only independent participants can recover the secret, our scheme's access structure is robust and reliable. Compared to other code-based schemes, the size of the secret and the number of coalitions of the proposed scheme is greater than the other schemes. So, the proposed scheme is more secure and efficient.

\end{document}